\documentclass[12pt,a4paper]{article}

\begin{document}
\begin{center}
{\large \bf Kustaanheimo-Stiefel transformation and static zero modes of
Dirac operator.}
\end{center}
\begin{center}
\large D.V. Aleynikov, E.A. Tolkachev
\end{center}
\begin{center}
{B.I. Stepanov Institute of Physics, National Academy of
Sciences of Belarus, \\ 68 F. Skaryna Avenue, 220072 Minsk,
Belarus\\ {\small \rm
   E-mail:     aleynik@dragon.bas-net.by \\ \hspace{1cm} tea@dragon.bas-net.by}}
\end{center}
\begin{center}
Abstract
\end{center}

{\small{By exploiting the relation between static zero modes of massless Dirac operator and
Kustaanheimo-Stiefel (Hopf) bundle sections, a general zero modes Ansatz which
depends on an arbitrary real vector-function on $R^3$ is constructed.}}

\vspace{1cm}
{\bf 1.} The problem of zero modes of the two, three and four-dimensional Dirac operator
in external gauge field is of interest in quantum mechanics as well as quantum
field theory. The zero modes affect strongly the ground state of spin 1/2 charged
particle in two-dimensional magnetic field, the stability or the collapse of
Coulomb systems with magnetic field and the behaviour of the Fermi determinant
in quantum field theory.

It is common knowledge \cite{Fry} the crucial role of degeneracy of zero modes in some physical
applications. The explicit examples of such a degeneracy have been constructed
in \cite{Nash}. But in \cite{Elton} were proposed two new examples of zero
modes in three dimensions.

In the present paper we deal with the three-dimensional static massless Dirac (Pauli) equation.
A general zero modes Ansatz  incorporating the previously known
solutions as special cases is proposed by using the quaternion form of expressions for  planar
$SO(3)$ rotations in the vector parameterization  \cite{Fedorov} as well as sections of
the Kustaanheimo-Stiefel (Hopf) bundle $\dot{R^4}\rightarrow \dot{R^3}$ ($S^3\rightarrow S^2$)
\cite{Pris1}, \cite{Pris2}.

{\bf 2.} For this purpose the linear biquaternion formulation of Dirac equation
\cite{Berezin} is used. In our approach all variables are biquaternions
(quaternions over the complex number field). Arbitrary biquaternion is
written as $q=q_0e_0+q_1e_1+q_2e_2+q_3e_3=q_0+\mathbf{q}$,
where $e_0, e_1, e_2$ and $e_3$ are basis elements.
The simplest realization of the basis is $I, -i\sigma_1, -i\sigma_2$ and $-i\sigma_3$,
where $I$ is the identity $2\times 2$ matrix, $\sigma_l$ are Pauli matrices.
Symbols $q^*$ and $\overline{q}$ denote the complex and quaternion
conjugations, respectively: $q^*=q_{0}^{*}+\mathbf{q}^*$, $\overline{q}=q_0-\mathbf{q}$.
Scalar part of biquaternion is $(q)_S=q_0$, vector one is $(q)_V=\mathbf{q}$.
The product of two biquaternions $a=a_0+\mathbf{a}$ and $b=b_0+\mathbf{b}$ is a
biquaternion $c=ab$, where
$(c)_S=c_0=a_0b_0-(\mathbf{a}\mathbf{b})$,
$(c)_V=\mathbf{c}=a_0\mathbf{b}+b_0\mathbf{a}+[\mathbf{a}\mathbf{b}]$;
$(\mathbf{a}\mathbf{b})=\delta_{lm}a_lb_m$ and
$[\mathbf{a}\mathbf{b}]=\varepsilon_{klm}a_lb_m$, $k,l,m= 1,2,3$.

There exists another useful biquaternion basis:
$$\Pi_{1(2)}=1/2(e_0\pm ie_3), \quad S_1=ie_1, \quad S_2=ie_2.$$
Projective quaternions $\Pi_{1(2)}$ (divisor of zero) and
quaternions $S_{1(2)}$ possess the following properties:
$$\Pi^{2}_{1(2)}=\Pi_{1(2)}, \quad \Pi_{1}\Pi_{2}=\Pi_{2}\Pi_{1}=0, \quad \Pi_{1}+\Pi_{2}=1,
\quad \overline{\Pi}_{1}=\Pi^{*}_{1}=\Pi_{2}.$$
$$\Pi_{1}S_{1(2)}=S_{1(2)}\Pi_{2}, \quad S_{1(2)}^{2}=1, \quad
S_{1(2)}^{*}=\overline{S}_{1(2)}=-S_{1(2)}.$$
Instead of $e_1, e_2$ and $e_3$ one
can choose, certainly, three arbitrary orthogonal unit
vectors $\mathbf{n}_1, \mathbf{n}_2$ and $\mathbf{n}_3,$ where
$(\mathbf{n}_l\mathbf{n}_m)=\delta_{lm}$.
Then the biquaternion form of Dirac equation \cite{Berezin} is
\begin{equation}
(\nabla_{(4)}+ieA)\Psi \Pi_{1}S_1+(\overline{\nabla}_{(4)}+ie\overline{A})\Psi \Pi_{2}S_1-m\Psi
=0,
\end{equation}
where $\Psi=2(\varphi_1+S_1\varphi_2)\Pi_1+2(\xi_1S_1+\xi_2)\Pi_2$,
$A=iA_0+\mathbf{A}$, $A^*=-\overline{A}$,
$\nabla_{(4)}=i\partial_{t}-\mathbf{\nabla}$, $\nabla_{(4)}^{*}=-\overline{\nabla}_{(4)}$.

For the massless static case the equation (1) can be divided in two equations
\begin{equation}
(\mathbf{\nabla}-ie\mathbf{A})\Psi \Pi_1=0, \quad
(\mathbf{\nabla}-ie\mathbf{A})\Psi \Pi_2=0,
\end{equation}
where $\Psi \Pi_1$ and $\Psi \Pi_2$ are realized by two ideals of biquaternion
algebra.
The equations (2) are equivalent to each other. For this reason it is
enough to solve, for instance, the first equation for biquaternion $\Psi \Pi_1$.

Let us take the basis as
$e_0, \mathbf{n}_1, \mathbf{n}_2$ and $\mathbf{n}_3$.
Then $\Pi_{1}=1/2(e_0+i\mathbf{n}_3)$,
$S_1=i\mathbf{n}_1$.
The first biquaternion equation (2) may be written now in terms of real quaternion.
Indeed, taking into account that  $i\Pi_1=-\mathbf{n}_3\Pi_1$ we have
\begin{eqnarray}
\Psi \Pi_1 & = & 2(\varphi_1+S_1\varphi_2)\Pi_1=2(\mathrm{Re}\varphi_1+i\mathrm{Im}\varphi_1-
\mathrm{Im}\varphi_2\mathbf{n}_1+i\mathrm{Re}\varphi_2\mathbf{n}_1)\Pi_1=
\nonumber \\
& =& 2(\mathrm{Re}\varphi_1-\mathrm{Im}\varphi_2\,  \mathbf{n}_1 +
\mathrm{Re}\varphi_2\, \mathbf{n}_2-\mathrm{Im}\, \varphi_1 \mathbf{n}_3)\Pi_{1}=2U\Pi_1.
\end{eqnarray}
Here $U$ is real quaternion.
Substituting  (3) into (2) we obtain the same equation for both the real and imaginary
parts of biquaternion:
\begin{equation}
\mathbf{\nabla}\{U\}\mathbf{n}_3-\mathbf{A}U=0,
\end{equation}
The quaternion derivative operator $\mathbf{\nabla}$ acts on the quaternion
being situated within the curly brackets only. From (4) we get
\begin{eqnarray}
\Bigl(\frac{\mathbf{\nabla}\{U\}\mathbf{n}_3\overline{U}}{U\overline{U}}\Bigr)_V=
\mathbf{A},  \nonumber \\
\Bigl(\frac{\mathbf{\nabla}\{U\}\mathbf{n}_3\overline{U}}{U\overline{U}}\Bigr)_S=0.
\end{eqnarray}

Using quaternion properties $(ab)_S=(ba)_S$, $(\overline{a})_S=(a)_S$ and
$(\overline{a})_V=-(a)_V$ gives
\begin{displaymath}
(\mathbf{\nabla}\{U\}\mathbf{n}_3\overline{U})_S=
(U\mathbf{n}_3\{\overline{U}\}\mathbf{\nabla})_S=
(U\mathbf{\nabla}\{\mathbf{n}_3\overline{U}\})_S=
1/2(\mathbf{\nabla}\{U\mathbf{n}_3\overline{U}\})_S=0.
\end{displaymath}
So, we have a single constraint on quaternion $U$, namely, vector-quaternion
$\mathbf{f}=U\mathbf{n}_3\overline{U}$ is a solenoidal one:
$(\mathbf{\nabla}\mathbf{f})_S=-(\mathbf{\nabla}\mathbf{f})=0$,
$\mathbf{f}=[\mathbf{\nabla}\mathbf{F}]$.

The expressions (5) are the quaternion form of those obtained in \cite{Loss}
(see also \cite{Elton}, proposition 1). There are many ways of dealing with these
equations. For instance, if
$(\mathbf{\nabla}\{U\}\mathbf{n}_3\overline{U})_S=
(\mathbf{\nabla}\{U\}\overline{U}\mathbf{n}_3)_S=0$ then from the second equation of (5)
we get
\begin{equation}
\mathbf{\nabla}\{U\}\overline{U}=\omega_1\mathbf{n}_1+\omega_2\mathbf{n}_2+\omega_3e_0,
\end{equation}
where $\omega_1$, $\omega_2$ and $\omega_3$ are arbitrary functions.
The simplest case corresponding to the choice $\omega_1=\omega_2=0$ was already
considered in \cite{Loss}. So,
\begin{equation}
\mathbf{\nabla}\{U\}=\omega U,
\end{equation}
($\omega=\omega_3/U\overline{U}$)
and the vector potential $\mathbf{A}$ takes form
$$\mathbf{A}=\omega
\frac{U\mathbf{n}_3\overline{U}}{U\overline{U}}.$$

{\bf 3.} However, our main goal is to build a general zero modes Ansatz without solving
equations (6) or (7). Indeed,  it would be quite enough to find $U(\mathbf{f})$ from
the algebraic equation
\begin{equation}
\mathbf{f}=U\mathbf{n}_3\overline{U}.
\end{equation}
It can be done in at least two rather similar ways.

The first one starts from the vector-quaternion expression for planar $SO(3)$ rotation of
$\mathbf{n}_3$ to $\mathbf{f}$ up to $SO(2)$ isotropy subgroup of $\mathbf{n}_3$
\cite{Fedorov}. The second way is based on the similarity of equation (8) and
Kustaanheimo-Stiefel bundle transformation $\dot{R^4}\rightarrow \dot{R^3}$
(the latter being isomorphic to Hopf one $S^3\rightarrow S^2$). The quaternion bundle
sections were constructed up to $U(1)$ fiber transformations in \cite{Pris1}.
In both the cases quaternion $U$ can be written as:
\begin{displaymath}
U=\sqrt{f}\frac{1+\mathbf{c}}{\sqrt{1+\mathbf{c}^2}},
\end{displaymath}
where vector-quaternion $\mathbf{c}$ is
\begin{displaymath}
\mathbf{c}=
\frac{[\mathbf{n}_3\, \mathbf{f}]}{f+(\mathbf{n}_3\mathbf{f})},
\end{displaymath}
or
\begin{equation}
U =\frac{\sqrt{f+(\mathbf{n}_3\mathbf{f})}}{\sqrt{2}}\biggl(1+
\frac{[\mathbf{n}_3\, \mathbf{f}]}{f+(\mathbf{n}_3\mathbf{f})}
\biggr).
\end{equation}
Substituting (9) into (5) we get vector-potential
\begin{equation}
\mathbf{A}=\frac{[\mathbf{\nabla}\mathbf{f}]}{2f}+
\frac{f-(\mathbf{n}_3\mathbf{f})}{2f}
\mathbf{\nabla}\{\mathbf{\hat{c}}\}\mathbf{\hat{c}}
\mathbf{n}_3.
\end{equation}
Here
\begin{displaymath}
f=\sqrt{\mathbf{f}^2}, \quad
\mathbf{\hat{c}}=
\frac{[\mathbf{n}_3\mathbf{f}]}{\sqrt{f^2-(\mathbf{n}_3\mathbf{f})^2}}, \quad
\mathbf{f}=[\mathbf{\nabla}\mathbf{F}].
\end{displaymath}

We note that transformations of isotropy subgroup of vector $\mathbf{n}_3$ or fiber in
the Kustaanheimo-Stiefel bundle describe the gauge freedom of quaternions $2U\Pi_1$ $(\Psi \Pi_1)$
and $\mathbf{A}$ satisfying  the first biquaternion static Pauli equation of (2):
$$\Psi \Pi_1 \rightarrow \exp(i\alpha)\Psi \Pi_1=2U\exp(-\alpha \mathbf{n}_3)\Pi_1=
2U\frac{1-\beta \mathbf{n}_3}{\sqrt{1+\beta^2}}\Pi_1 \, ,
\quad \alpha=\arctan \beta \, ,$$
$$\mathbf{A}\rightarrow \mathbf{A}+\mathbf{\nabla}\{\arctan \beta\}.$$

The general solutions (9) and (10) of static Pauli equation
\begin{displaymath}
\sigma_l(\partial_l-iA_l)\psi=0
\end{displaymath}
can be easily written in spinor and vector notations:
\begin{eqnarray}
\psi & = & \frac{1}{\sqrt{2(f+(\mathbf{f}\mathbf{n}_3))}}{f+(\mathbf{f}\mathbf{n}_3)
\choose (\mathbf{f}\mathbf{n}_1)+i(\mathbf{f}\mathbf{n}_2)},  \\
\mathbf{A} & = & \frac{\mathrm{curl}\ \mathbf{f}}{2f}+\frac{f-(\mathbf{f}\mathbf{n}_3)}{2f}
\mathrm{grad}\Bigl(\arctan
\frac{(\mathbf{f}\mathbf{n}_2)}{(\mathbf{f}\mathbf{n}_1)}\Bigr), \\
\mathbf{f} & = & \mathrm{curl}\mathbf{F}, \nonumber
\end{eqnarray}
where $\mathbf{F}$ is an arbitrary vector-function.

The solutions  (11) and (12) reproduce well-known
special cases. For instance, let
\begin{displaymath}
\mathbf{F}=\frac{1}{4(1+r^2)^2}(2[\mathbf{n}_3\mathbf{r}]+
2(\mathbf{n}_3\mathbf{r})\mathbf{r}+(1-r^2)\mathbf{n}_3).
\end{displaymath}
Then the spinor is
\begin{displaymath}
\psi=\frac{1-i(\mathbf{n}_3\mathbf{r})}{\sqrt{1+(\mathbf{n}_3\mathbf{r})^2}}
\frac{1}{(1+r^2)^{3/2}}{1+i(\mathbf{n}_3\mathbf{r})
\choose -(\mathbf{n}_2\mathbf{r})+i(\mathbf{n}_1\mathbf{r})}
\end{displaymath}
and vector-potential is
\begin{displaymath}
\mathbf{A}=12\mathbf{F}-\mathbf{\nabla}\{\arctan(\mathbf{n}_3\mathbf{r})\}.
\end{displaymath}
By choosing $\mathbf{n}_3$ along the $z$ axis, i.e., $\mathbf{n}_3=(0,0,1)$ we get
the solution \cite{Nash} up to the gauge transformation mentioned above,
\begin{displaymath}
\psi=\frac{1}{(1+r^2)^{3/2}}{1+iz \choose -y+ix},
\end{displaymath}
\begin{displaymath}
\mathbf{A}=\frac{3}{(1+r^2)^2}(2xz-2y,\, 2yz+2x,\,
z^2+1-x^2-y^2).
\end{displaymath}

The next example having the monopole-like structure is
\begin{displaymath}
\mathbf{F}=\frac{1}{r}
\frac{[\mathbf{n}_3\mathbf{r}]}{r+(\mathbf{n}_3\
\mathbf{r})}\, .
\end{displaymath}
Substituting this equation into (9) and (10) we obtain quaternion function
\begin{displaymath}
U=\frac{\sqrt{r+(\mathbf{n}_3\mathbf{r})}}{r\sqrt{2r}}\biggl(1+
\frac{[\mathbf{n}_3\mathbf{r}]}{r+(\mathbf{n}_3\mathbf{r})}
\biggr),
\end{displaymath}
spinor
\begin{displaymath}
\psi=\frac{1}{r\sqrt{2r(r+(\mathbf{n}_3\mathbf{r}))}}
{r+(\mathbf{n}_3\mathbf{r}) \choose (\mathbf{n}_1\mathbf{r})+i(\mathbf{n}_2\mathbf{r})}
\end{displaymath}
and vector-potential
\begin{displaymath}
\mathbf{A}=\frac{1}{2r}
\frac{[\mathbf{n}_3\mathbf{r}]}{r+(\mathbf{n}_3\
\mathbf{r})}\, .
\end{displaymath}
The choice of $\mathbf{n}_3$ along the $z$ axis  leads to the solution obtained in \cite{Freund}.

It should be emphasized that monopole potential is a connection on the
Kustaanheimo-Stiefel bundle. Moreover, quaternion $U$ includes a monopole-like term.
So, this potential appears both as a connection and a section of bundle.
This fact was investigated in more detail in \cite{Pris1}.

In addition, we note
that if $\sqrt{|\mathrm{curl}\mathbf{F}|}$ is a square-integrable function, i.e.,
$\sqrt{|\mathrm{curl}\mathbf{F}|}\in L^2$ then so is the spinor $\psi$. The
latter fact follows from equation $|\psi|^2=|\mathrm{curl}\mathbf{F}|$.

We note in conclusion the examples of zero modes proposed in \cite{Elton} can
be also represented in form (11) and (12). The corresponding vector-function
$\mathbf{F}$ depends on two scalar function connected by subsidiary
conditions.

The authors are grateful to dr. E.V. Doktorov for fruitful discussions.

\end{document}